\documentclass[11pt,a4paper,english,superscriptaddress]{revtex4}
\usepackage{amssymb}
\usepackage{amsmath} 
\usepackage{slashed}
\usepackage{graphicx}
\usepackage{babel}
\usepackage{hyperref}
\usepackage{color}
\usepackage{comment}
\makeatletter\AtBeginDocument{\let\@elt\relax}\makeatother

\begin{document}

\title{Universality of gauge coupling constant in the Einstein-QED system}

\author{L.~Ibiapina~Bevilaqua}
\email{leandro@ect.ufrn.br}
\affiliation{Escola de Ci\^encias e Tecnologia, Universidade Federal do Rio Grande do Norte\\
Caixa Postal 1524, 59072-970, Natal, Rio Grande do Norte, Brazil.}

\author{A.~C.~Lehum}
\email{lehum@ufpa.br}
\affiliation{Faculdade de F\'isica, Universidade Federal do Par\'a, 66075-110, Bel\'em, Par\'a, Brazil.}

\author{Huan Souza}
\email{huan.souza@icen.ufpa.br}
\affiliation{Faculdade de F\'isica, Universidade Federal do Par\'a, 66075-110, Bel\'em, Par\'a, Brazil.}

\begin{abstract}
In this paper we discuss the universality of the renormalization of the gauge coupling constant in the quantum electrodynamics coupled to the Einstein's gravity in the framework of effective field theory in an arbitrary gauge. We observe that the renormalization of the three-point functions with different particles receive different contributions from the gravitational sector. Despite that, the universality of the gauge coupling constants is guaranteed by the Ward identity that makes the renormalized electric charge receive no contribution from the gravitational sector at one-loop order.  
\end{abstract}


\maketitle

\section{Introduction}

Although the Standard Model is a very successful description of the fundamental interactions of matter, it is also notoriously incomplete. In particular, it does not include the most easily perceived interaction of all: gravity. 

Unlike the gauge theories that describe the electroweak and the strong interactions in the Standard Model, the quantization of Einstein's general relativity leads to a nonrenormalizable quantum field theory~\cite{'tHooft:1974bx,PhysRevLett.32.245,Deser:1974cy}. Nevertheless, it is perfectly possible to include gravity in the quantum realm, if only we agree to restrict ourselves to low energies compared to the Planck scale. In the effective field theory (EFT) of gravity we have a consistent and reliable way to compute quantum gravitational corrections to physical quantities~\cite{Donoghue:1994dn} and the potential harm of nonrenormalizability is tamed \cite{Donoghue:2017pgk,Burgess:2003}. 

This EFT approach has been used to explore the role of gravity in many aspects of the Standard Model. One very interesting example is the claim done in Ref.~\cite{Chang:2012zzo} that the presence of gravity may have a nontrivial effect in the renormalization of theories with spontaneous symmetry breaking. Moreover, it has been argued that exact continuous global symmetries cannot exist in a quantum field theory including gravity~\cite{Banks:2010zn,Harlow:2018tng,Harlow:2018jwu}, and the authors of Ref.~\cite{Fichet:2019ugl} suggest that global symmetries can be realized in an EFT including gravity only as an approximation, i.e., as an emergent symmetry. In addition, there was an intense debate over the gravitational contribution to the beta function of gauge theories~\cite{Robinson:2005fj,Pietrykowski:2006xy,Felipe:2012vq,Ebert:2007gf,Nielsen:2012fm,Toms:2008dq,Toms:2010vy,Ellis:2010rw,Anber:2010uj,Bevilaqua:2015hma}.

In \cite{Anber:2010uj} it was argued that the definition of running coupling constants for effective field theories of nonrenormalizable theories such as gravity are much trickier than in renormalizable theories. That is because the quantum corrections in a nonrenormalizable model are associated with the renormalization of higher order operators rather than the original action. Although it is possible to define a running coupling within a given process, the universality of the quantum effects is not guaranteed.

In this work, we use dimensional regularization to study the running of the coupling constant of QED with two massive fermions. Unlike most of the previous work on this subject, our analysis is not based on the calculation of the effective potential and we use an arbitrary gauge to renormalize the n-point functions the theory. We show that the renormalization of the vertex functions are process dependent, unlike QED in the absence of gravity. However, our results show that the universality of the gauge coupling constant is not affected by gravity. We also observed that the Ward identity holds in the presence of gravity.

Throughout this paper we use natural units $c=\hbar=1$.

\section{The effective field theory approach}\label{sec11}

Renormalization is the procedure to redefine the parameters and fields of the Lagrangian in such a way that the divergencies in the loop corrections to the amplitudes are exactly canceled. This is possible only in theories where all the divergencies can be traced to terms present in the Lagrangian.

Gravity is notoriously nonrenormalizable, meaning that the divergencies demand redefinitions of infinitely many parameters. This is because the divergencies generated by loop corrections correspond to terms of a higher mass dimension than those in the bare Lagrangian, demanding new terms. When we add the required terms to the Lagrangian, other divergencies appear demanding additional terms in an endless cycle. 

The idea behind the effective field theory (EFT) approach is to organize the (infinitely many) extra terms required by the divergencies according to the dimension of the coupling. This procedure will describe the low-energy behavior of the theory. The additional higher-order terms describe the (indirect) effects of degrees of freedom that are ``active'' only at higher energies.

In order to better understand the construction of the EFT, let us consider a field theory with characteristic energy scale $E_0$ and suppose there is a separation of scales, meaning that it is possible to choose a cutoff $\Lambda$ at or slightly below $E_0$ such that we can divide the fields as
\[
\phi = \phi_H + \phi_L, \qquad
\left\{
\begin{array}{l}
\phi_H: \omega > \Lambda \\
\phi_L: \omega < \Lambda
\end{array}
\right.
\]

For example, in a model with complex scalar $\Phi \rightarrow e^{i\alpha} \Phi$ described by
\begin{equation}\label{u1}
{\cal L} = - |\partial_\mu\Phi|^2 - \lambda (|\Phi|^2 - v^2)^2
\end{equation}
where the potential has a minimum at $\langle\Phi\rangle = v$, we have a separation of scales. This can be easily seen introducing the real fields $\theta(x)$ and $h(x)$ through \begin{equation}
\Phi(x) = \Big(v+h(x)\Big)e^{i\theta(x)}
\end{equation}
and then writing \eqref{u1} as
\begin{equation}\label{u1-exp}
{\cal L} = - (\partial h)^2 - M^2 h^2 - v^2 (\partial\theta)^2 - 4v\lambda h^3 - \lambda h^4 -2vh(\partial\theta)^2 - h^2(\partial\theta)^2,
\end{equation}
with $M^2 \equiv 4\lambda v^2$. From the above Lagrangian, we can see that the (massless) Goldstone boson $\theta(x)$ is the light mode, while the heavy mode is the massive field $h(x)$. That means we can use EFT logic to write an effective theory for the Goldstone $\theta(x)$ valid for scales much less than $M$, the mass of the heavy field $h(x)$.

As we identify the separation of scales, we can integrate the heavy fields $\phi_H$ in the functional of the theory,
\[
\int{\cal D} \phi_L{\cal D}\phi_H e^{iS[\phi_L,\phi_H]} = \int{\cal D} \phi_L e^{iS_{\Lambda}[\phi_L]},
\]
and thus define an (``effective'') action $S_{\Lambda}[\phi_L]$ that depends explicitly only on $\phi_L$, but includes (hidden) the degrees of freedom of $\phi_H$. This effective action describes the low-energy dynamics of the light fields $\phi_L$, where $\phi_H$ will never appear as external particles in any scattering due to the high energetic cost to produce it. 

Of course, due to quantum fluctuations, the heavy field $\phi_H$ will manifest itself in the internal lines and its effects must be somehow encoded in $S_{\Lambda}$. Thus, for example, if we have in the complete theory the diagram
\begin{center}
\includegraphics[scale=0.1]{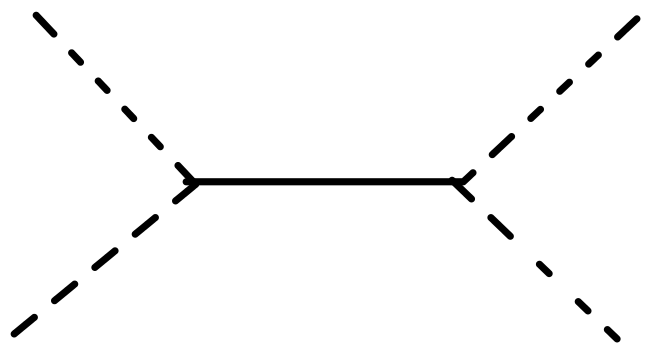} \qquad\qquad ($\phi_L= dashed$ and $\phi_H=full$),
\end{center}
in the effective theory, this diagram is not allowed because we do not have the heavy fields in the action and the diagrams should be composed only by dashed lines, so that diagram must be effectively replaced by
\begin{center}
\includegraphics[scale=0.2]{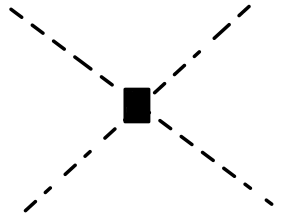}
\end{center}
with the appropriate Feynman rule.

When we approximate internal $\phi_H$ lines by a constant, what we are doing is expanding the heavy particle propagator:
\begin{equation}
\frac{i}{p^2-M^2} \rightarrow -\frac{i}{M^2}\left( 1 + \frac{p^2}{M^2} + \cdots \right)
\end{equation}
and using the fact that $M$ is much bigger than the typical energy of the process ($p^2 \ll M^2$) to truncate the series. So, the heavy particle, in first approximation, can be replaced by an interaction term in the Lagrangian (with a derivative coupling to account for the momentum dependence).

When the complete theory is not known (as in the case of gravity), we can nevertheless guess the effective action guided by the symmetries to construct $S_\Lambda$. Although symmetry considerations allow us to write an infinite number or terms, they can be organized as an energy expansion. Thus, a term of $S_\Lambda$ can be written as
\[
S^{(i)}_\Lambda = \Lambda^{D-\delta_i}\int d^D x \lambda_i {\cal O}_i
\]
where $\delta_i$ is the mass dimension of the operator ${\cal O}_i$, the coupling $\lambda_i$ is dimensionless and $\Lambda$ is an energy scale. Thus, as we go to terms with higher mass dimension, we have terms weighed with higher powers of $(1/\Lambda )$ and thus less relevant.

The EFT to describe the low energy dynamics of \eqref{u1}, for example, can be guessed without bothering to integrate out the $h(x)$ field from \eqref{u1-exp}. Considering only that the original theory has an $U(1)$ symmetry (corresponding to a shift: $\theta \rightarrow \theta + a$), we know that the allowed terms contain derivatives of $\theta(x)$. Considering also the Lorentz invariance and parity ($\theta \rightarrow - \theta$), we then write the Lagrangian ``organized as an energy ({\it i.e.} derivative) expansion":
\begin{equation}\label{eft1}
{\cal L}_{eff} = -\frac{1}{2}(\partial\theta)^2 + \frac{g_1}{\Lambda^4}(\partial\theta)^4 +
\frac{g_4}{\Lambda^6} (\partial_\mu\partial_\nu\theta\partial^\mu\partial^\nu\theta) (\partial\theta)^2
+ \frac{g_2}{\Lambda^8}(\partial\theta)^6 + \frac{g_3}{\Lambda^6} (\Box \theta)^2 (\partial\theta)^2  + \cdots
\end{equation}
such that the parameters $g_i$ are dimensionless and $\Box=\partial^\mu \partial_\mu$ is the d’Alembert operator.

This procedure will describe the low-energy behavior of the theory even if some other degrees of freedom were to be manifested in higher energies, since we are including all terms allowed by the symmetry.

In a nonrenormalizable field theory, as we add more and more terms to absorb the divergencies, we are adding less and less relevant terms. Moreover, the symmetries of the theory will ensure that at any given dimension, only a finite number of terms is allowed. Thus, if we agree to stay at the low energy regime of the nonrenormalizable theory, it is possible to have a well-defined effective theory adding only a finite number of extra terms in the Lagrangian.

\section{The effective field theory of quantum gravity coupled to QED.}\label{sec2}

We start with the effective Lagrangian describing two massive fermions interacting via gravity and an Abelian gauge field:
\begin{eqnarray}\label{fQED}
\mathcal{L}=&& \sqrt{-g}\sum_l\Big\{\frac{2}{\kappa^2}R-\frac{1}{4} g^{\mu\nu}g^{\alpha\beta} F_{\mu\nu} F_{\alpha\beta} +i\bar{\psi}_l(\nabla_{\mu} - ieA_{\mu})\gamma^{\mu}\psi_l - m_l\bar{\psi}_l\psi_l \nonumber \\
&&\qquad\qquad +\mathcal{L}_{HO} +\mathcal{L}_{GF}+\mathcal{L}_{CT}\Big\},
\end{eqnarray}
\noindent where the index $l$ runs over the generation of leptons (only electron and muon for simplicity) and we wrote the Dirac matrices contracted with the vierbein ($\gamma^{\mu} \equiv \gamma^a e^{\mu}_a$), $\displaystyle{\kappa^2=32\pi G=\frac{32\pi}{M_P^2}}$, with $M_P$ being the Planck mass and $G$ the Newtonian gravitational constant. Moreover, we have indicated above the extra terms needed for the renormalization of the model: $\mathcal{L}_{GF}$ is the gauge-fixing plus Faddeev-Popov ghost Lagrangian (for both graviton and photon), $\mathcal{L}_{CT}$ is the Lagrangian of counterterms and $\mathcal{L}_{HO}$ is the Lagrangian of higher derivative terms, where the terms relevant to our analysis are shown below:
\begin{eqnarray}\label{ho}
\mathcal{L}_{HO}=i\bar\psi_l~\frac{\Box}{M_P^2}\left(\tilde{e}_1\slashed{\partial}- \tilde{e}_2 m_l\right)\psi_l-\frac{\tilde{e}_3}{4}F^{\mu\nu}\frac{\Box}{M_P^2} F_{\mu\nu}+\frac{i\tilde{e}_4}{M_P^2}\bar\psi_l\gamma^\mu\partial^\nu\psi_l F_{\mu\nu}+ \frac{\tilde{e}_5}{M_P^2}(\bar{\psi}_l\gamma^\mu\psi_l)^2\cdots,
\end{eqnarray}
\noindent where 
\begin{equation}
 {(\bar{\psi}_l\gamma^\mu\psi_l)^2 = (\bar{\psi}_1\gamma^\mu\psi_1)(\bar{\psi}_1\gamma_\mu\psi_1) + 2(\bar{\psi}_1\gamma^\mu\psi_1)(\bar{\psi}_2\gamma_\mu\psi_2) + (\bar{\psi}_2\gamma^\mu\psi_2)(\bar{\psi}_2\gamma_\mu\psi_2),}
\end{equation}
  are the four-fermion operators and $\tilde{e}_i$ are dimensionless couplings.

In the above expression for $\mathcal{L}_{HO}$, $\tilde{e}_i$ are dimensionful coupling constants and the indices are raised and lowered with the flat metric $\eta_{\mu\nu}=(+,-,-,-)$ and $\Box =\eta^{\mu\nu}\partial_\mu\partial_\nu$, since in order to work with \eqref{fQED} we have to first expand $g_{\mu\nu}$ around the flat metric
\[
g_{\mu\nu} = \eta_{\mu\nu} + \kappa h_{\mu\nu}.
\]

The resulting Lagrangian is given by
\begin{eqnarray}\label{L}
 \mathcal{L} = \mathcal{L}_h + \mathcal{L}_f + \mathcal{L}_A +\mathcal{L}_{HO} + \mathcal{L}_{GF} + \mathcal{L}_{CT},
\end{eqnarray}
\noindent where~\cite{Choi:1994ax}
\begin{subequations}\label{Lexp}
 \begin{eqnarray}
  &&\mathcal{L}_h = -\frac{1}{4}\partial_\mu h\partial^\mu h +\frac{1}{2}\partial_{\mu}h^{\sigma\nu}\partial^\mu h_{\sigma\nu};\label{Lh}\\
  &&\mathcal{L}_f = \frac{i}{2}\left(\bar{\psi}_l\gamma^\mu\partial_\mu\psi_l - \partial_\mu\bar{\psi}_l\gamma^\mu\psi_l\right) - m\bar{\psi}_l\psi_l + e\bar{\psi}_l\gamma^\mu\psi_l A_\mu\nonumber\\
  &&\quad\quad~+\frac{\kappa}{2}h\left[\frac{i}{2}(\bar{\psi}_l\gamma^\mu\partial_\mu\psi_l - \partial_\mu\bar{\psi}_l\gamma^\mu\psi_l) - m\bar{\psi}_l\psi_l\right] - \frac{\kappa}{4}h_{\mu\nu}\left(\bar{\psi}_l\gamma^\mu\partial\nu\psi_l - \partial^\nu\bar{\psi}_l\gamma^\mu\psi_l\right)\nonumber\\
  &&\quad\quad~-\frac{1}{2}\kappa e\left(h\eta_{\mu\nu}-h_{\mu\nu}\right)\bar{\psi}_l\gamma^\mu\psi_l A^\nu;\label{Lf}\\
  &&\mathcal{L}_A = -\frac{1}{4}F^{\mu\nu}F_{\mu\nu} + \frac{\kappa}{2}h^\tau_{~\nu}F^{\mu\nu}F_{\mu\tau} - \frac{\kappa}{8}hF^{\mu\nu}F_{\mu\nu};\label{LA}\\
  &&\mathcal{L}_{GF}=\frac{1}{\xi_h}\left(\partial^\nu h_{\mu\nu}-\frac{1}{2}\partial_\mu h\right)^2 - \frac{1}{2\xi_A}(\partial_\mu A^\mu)^2\label{LGF},
 \end{eqnarray}
\end{subequations}
\noindent where $h=h^\mu_{~\mu}, F^{\mu\nu}$ is the usual electromagnetic field strength, $\xi_h$ is the gravitational gauge-fixing parameter, and $\xi_A$ is the electromagnetic gauge-fixing parameter. This model was implemented through a set of \textit{Mathematica}\textsuperscript{TM} packages \cite{feyncalc,feynarts,feynrules,feynhelpers,FormCalc}. More details can be found online in Ref. \cite{site_lehum}.

Since the original Lagrangian in \eqref{fQED} is invariant under the (gravitational) gauge transformations
\begin{subequations}
 \begin{eqnarray}
 &&x^\mu\rightarrow x'^\mu=x^\mu + \kappa\xi^\mu(x);\\
 &&g^{\mu\nu}\rightarrow g'^{\mu\nu} = g^{\mu\nu} + \partial^\mu\xi^\nu + \partial^\nu\xi^\mu;\\
 &&\partial^\mu \rightarrow \partial'^\mu = \partial^\mu - (\partial^\mu\xi_\nu)\partial^\nu;\\
 &&\psi\rightarrow\psi'(x')=\psi(x),
 \end{eqnarray}
\end{subequations}
this expanded Lagrangian in \eqref{Lexp} is expected to be invariant under the linearized gauge transformation, with the graviton field transforming as
\begin{eqnarray}
 &&h^{\mu\nu}\rightarrow h'^{\mu\nu} = h^{\mu\nu} - \partial^\mu\xi^\nu - \partial^\nu\xi^\mu +\mathcal{O}(\kappa),
\end{eqnarray}
up to the order we are considering the expansion.

The electric charge is defined  as the coupling constant of the $\bar{\psi}\gamma^\mu\psi A_\mu$ vertex. However, due to the gauge invariance, we can compute the renormalized electric charge by renormalizing any of the terms in the expanded Lagrangian \eqref{Lexp} which contains it. In other words, although we are going to compute the renormalized electric charge from the vertex $e\bar{\psi}\gamma^\mu\psi A_\mu$ in this paper, we could also have computed it from the vertex $\frac{e\kappa}{2}(h\eta_{\mu\nu}-h_{\mu\nu})\bar{\psi}\gamma^\mu\psi A^\nu$. We will return to this point in Sec.\ref{sec4}

\begin{figure}[h!]
	\includegraphics[angle=0 ,width=8cm]{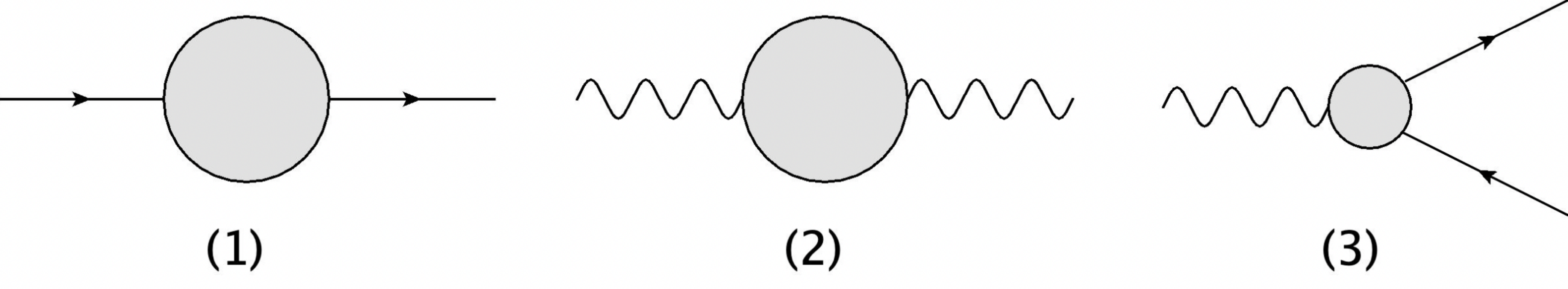}
	\caption{Feynman diagrams in QED. The renormalization of the model is achieved imposing certain conditions over them.}
	\label{fig01}
\end{figure}

Lastly, in order to compute the two-point and vertex functions, Fig. \ref{fig01}, we also have to discuss the propagators of the model. The quadratic part of the action photon field is given by the sum of the Maxwell's term with the high derivative invariant $-\frac{\tilde{e}_3}{4}F^{\mu\nu}\frac{\Box}{M_P^2} F_{\mu\nu}$,
\begin{eqnarray}
 S_{photon}&=&\int{d^4x}\Bigg[-\frac{1}{4}F^{\mu\nu}\left( 1+\tilde{e}_3\frac{\Box}{M_P^2}\right) F_{\mu\nu}-\frac{1}{2\xi_A}\left(\partial^\mu A_\mu\right)^2 \Bigg]\nonumber\\
 &=&\frac{1}{2}\int{d^4x}A_{\mu}\Bigg[\eta^{\mu\nu}\Box-\left(1-\frac{1}{\xi_A}\right)\partial^\mu \partial^\nu+\tilde{e}_3\frac{\Box}{M_P^2}\left(\eta^{\mu\nu}\Box-\partial^\mu \partial^\nu \right) \Bigg]A_{\nu},
 \end{eqnarray}
\noindent from which we find the following propagator in momentum space
\begin{eqnarray}\label{prop_ho1}
\Delta^{\mu\nu}(p) &=& -\frac{i}{p^2(1+\tilde{e}_3~p^2/M_P^2)}\left[\eta^{\mu\nu}-\left(1-\xi_A\left(1+\tilde{e}_3~\frac{p^2}{M_P^2}\right)\right)\frac{p^\mu p^\nu}{p^2} \right]\nonumber\\
&=& -i\left(\frac{1}{p^2}-\frac{\tilde{e}_3}{\tilde{e}_3~p^2+M_P^2}\right)\left[\eta^{\mu\nu}-\frac{p^\mu p^\nu}{p^2}+\xi_A\left(1+\tilde{e}_3~\frac{p^2}{M_P^2}\right)\frac{p^\mu p^\nu}{p^2} \right].
\end{eqnarray}

In fact, the effect of gravitational field on the photon propagator comes from the high derivative invariant $-\frac{\tilde{e}_3}{4}F^{\mu\nu}\frac{\Box}{M_P^2} F_{\mu\nu}$, but not only this term, the complete propagator should involve higher and higher derivative terms. But as we discussed before, the EFT should give precise descriptions for processes below the natural cutoff of the theory, i.e., the Planck scale. Therefore, we can expand the propagator (\ref{prop_ho1}) for $p^2\ll M_P^2$,   
\begin{eqnarray}\label{prop_ho1}
\Delta^{\mu\nu}(p) &=& -\frac{i}{p^2}\left[1+\tilde{e}_3\frac{p^2}{M_P^2}\left(1+\mathcal{O}(p^2/M_P^2)\right)\right]\left(\eta^{\mu\nu}-\frac{p^\mu p^\nu}{p^2} \right),
\end{eqnarray}
\noindent where we have chosen the Landau gauge $\xi_A=0$ in the last step in order to simplify and focus on the physical content of the series expansion. Notice that for small enough momenta, the gravitational corrections to the photon propagator are strongly suppressed by the Plank scale $M_P$. This feature allows us to neglect the high derivative terms for the propagator calculation, keeping in mind that the perturbative validity of our EFT is below the Planck scale.

On the other hand, the dominant behavior of photon propagator for momenta scales around Planck mass is a constant,
\begin{eqnarray}\label{prop_ho2}
\Delta^{\mu\nu}(p) &=& -\frac{i}{M_P^2(1+\tilde{e}_3)}\left[1-\frac{1+2\tilde{e}_3}{1+\tilde{e}_3}\frac{(p^2-M_P^2)}{M_P^2}+\mathcal{O}(p^2-M_P^2)^2\right]\left(\eta^{\mu\nu}-\frac{p^\mu p^\nu}{p^2} \right),
\end{eqnarray}
\noindent i.e., the electromagnetic interaction behaves like a point interaction. 

The high derivative terms in Lagrangian become dominant if we extrapolate the validity of our EFT, i.e., expanding (\ref{prop_ho1}) for $p^2\gg M_P^2$ we have     
\begin{eqnarray}\label{prop_ho3}
\Delta^{\mu\nu}(p) &=& -i\frac{M_P^2/\tilde{e}_3}{p^4}\left[1-\frac{M_P^2/\tilde{e}_3}{p^2}+\mathcal{O}(M_P^2/p^2)^2\right]\left(\eta^{\mu\nu}-\frac{p^\mu p^\nu}{p^2} \right).
\end{eqnarray}
\noindent Of course, this is an extrapolation of the validity of our EFT where the perturbative character of the expansion like (\ref{eft1}) is no longer valid.

So, from now on, considering processes below Planck scale, we use the following propagators in arbitrary gauge 
\begin{subequations}
\begin{eqnarray}
S_F(p) &=& i\frac{\slashed{p}+m_l}{p^2-m_l^2};\\
\Delta^{\mu\nu}(p) &=& -\frac{i}{p^2}\left(\eta^{\mu\nu}-(1-\xi_A)\frac{p^\mu p^\nu}{p^2} \right);\\
\Delta^{\alpha\beta\mu\nu}(p) &=& \frac{i}{p^2}\left(P^{\alpha\beta\mu\nu}-(1-\xi_h)\frac{Q^{\alpha\beta\mu\nu}}{p^2}\right),  
\end{eqnarray}
\end{subequations}
\noindent where $S_F(p)$, $\Delta^{\mu\nu}(p)$ and $\Delta^{\alpha\beta\mu\nu}(p) $ are the leptons, photon and graviton propagators, respectively. The projectors $P^{\alpha\beta\mu\nu}$ and $Q^{\alpha\beta\mu\nu} $ are given by
\begin{eqnarray}
P^{\alpha\beta\mu\nu} &=&\frac{1}{2} \left(\eta^{\alpha\mu}\eta^{\beta\nu}+\eta^{\alpha\nu}\eta^{\beta\mu}-\eta^{\alpha\beta}\eta^{\mu\nu} \right);\nonumber\\
Q^{\alpha\beta\mu\nu} &=& (\eta^{\alpha\mu}p^\beta p^\nu+\eta^{\alpha\nu}p^\beta p^\mu+\eta^{\beta\mu}p^\alpha p^\nu+\eta^{\beta\nu}p^\alpha p^\mu).
\end{eqnarray}

\section{Renormalization conditions}\label{sec3}

Renormalization is the procedure to relate the free (bare) parameters of the quantum field theory with their experimentally measured (renormalized) values. This is achieved imposing finiteness to the Feynman amplitudes.

We start redefining the fields as $A^\mu\rightarrow Z_3^{1/2}A^\mu$ and $\psi_l\rightarrow Z_{2l}^{1/2}\psi_l$, where $Z$ are the renormalizing functions, organized as a perturbative series
\begin{eqnarray}
Z=Z^{(0)}+ Z^{(1)}+Z^{(2)}+\cdots, \qquad \text{with} \quad Z^{(0)} = 1.
\end{eqnarray}
We omit here the redefinition of the gravitational field $h^{\mu\nu}$ because we are interested only in the gravitational correction to the QED sector (see Fig.\ref{fig01}). The relation between bare and renormalized electric charge is written in terms of the $Z$ functions and is given by
\begin{eqnarray}\label{eq_e_0}
e_0&=&\mu^{2\epsilon}\frac{Z_1}{Z_2 Z_{3}^{1/2}}e,
\end{eqnarray}
\noindent where $\mu$ is a mass scale introduced by the dimensional regularization, used to regularize the UV divergences in the Feynman amplitudes, and $\epsilon$ is related to the spacetime dimension $D$ by $D=4-2\epsilon$. Again we do not bother to write down the corresponding relation for the gravitational coupling constant $\kappa$ since it is not the subject of our analysis.

The relevant diagrams for us are the leptons and photon self-energies (respectively $\Sigma_l$ and $\Pi$) and the vertex function $\Gamma^\mu$. As in standard QED, we impose the following renormalization conditions:
\begin{subequations}\label{ren_cond_qed}
\begin{eqnarray}
\Sigma_l(\slashed{p})\Big{|}_{\slashed{p}=m_l}=0;\label{ren_cond_qed1}\\
\frac{d}{d\slashed{p}}\Sigma_l(\slashed{p})\Big{|}_{\slashed{p}=m_l}=0;\label{ren_cond_qed2}\\
\Pi(p^2)\Big{|}_{p^2=M^2}=0;\label{ren_cond_qed3}\\
-ie\Gamma^{\mu}(p1,p2,p3)\Big{|}_{(p_2-p_3=0)}=-ie\gamma^\mu,\label{ren_cond_qed4}
\end{eqnarray}
\end{subequations}      
where $\slashed{p}=\gamma^\mu p_\mu$ in the contraction of the Dirac gamma matrices with the four-momentum $p_\mu$. The first condition, Eq.(\ref{ren_cond_qed1}), fixes the mass of the lepton $l$ (electron or muon), while Eq.(\ref{ren_cond_qed2}) fixes the residue of the lepton propagators to be $1$. For the condition expressed in Eq.(\ref{ren_cond_qed3}), we chose to fix the pole of the photon propagator at renormalization scale $M^2$ only because this renormalization point will be convenient for the additional conditions we will impose below for the derivative of $\Pi (p^2)$, where the on-shell renormalization point ({\it i.e.,} at $p^2=0$) would be troublesome. Finally, Eq.(\ref{ren_cond_qed4}) fixes the electric charge. 

In addition to Eq.(\ref{ren_cond_qed}), we have to impose another set of four renormalization conditions to fix the renormalization of the high order operators, Eq.(\ref{ho}). One possibility, similar to the standard QED, is 
\begin{subequations}\label{ren_cond_ho}
\begin{eqnarray}
\frac{d}{dp^2}\Sigma_l(\slashed{p})\Big{|}_{\slashed{p}=m_l}=0;\label{ren_cond_ho1}\\
\frac{d}{dp^2}\frac{d}{d\slashed{p}}\Sigma_l(\slashed{p})\Big{|}_{\slashed{p}=m_l}=0;\label{ren_cond_ho2}\\
\frac{d}{dp^2}\Pi(p^2)\Big{|}_{p^2=M^2}=0;\label{ren_cond_ho3}\\
\frac{d\Gamma^{\mu}(p_1,p_2,p_3)}{d\slashed{p_1}}\Big{|}_{(p_2-p_3=0)}=0,\label{ren_cond_ho4}
\end{eqnarray}
\end{subequations}
\noindent where (\ref{ren_cond_ho4}) expresses the convenient choice to make $\tilde{e}_4$ to vanish at the renormalization scale. 

\section{Gravitational correction to the electric charge}\label{sec4}

\begin{figure}[h]
	\includegraphics[angle=0 ,width=6cm]{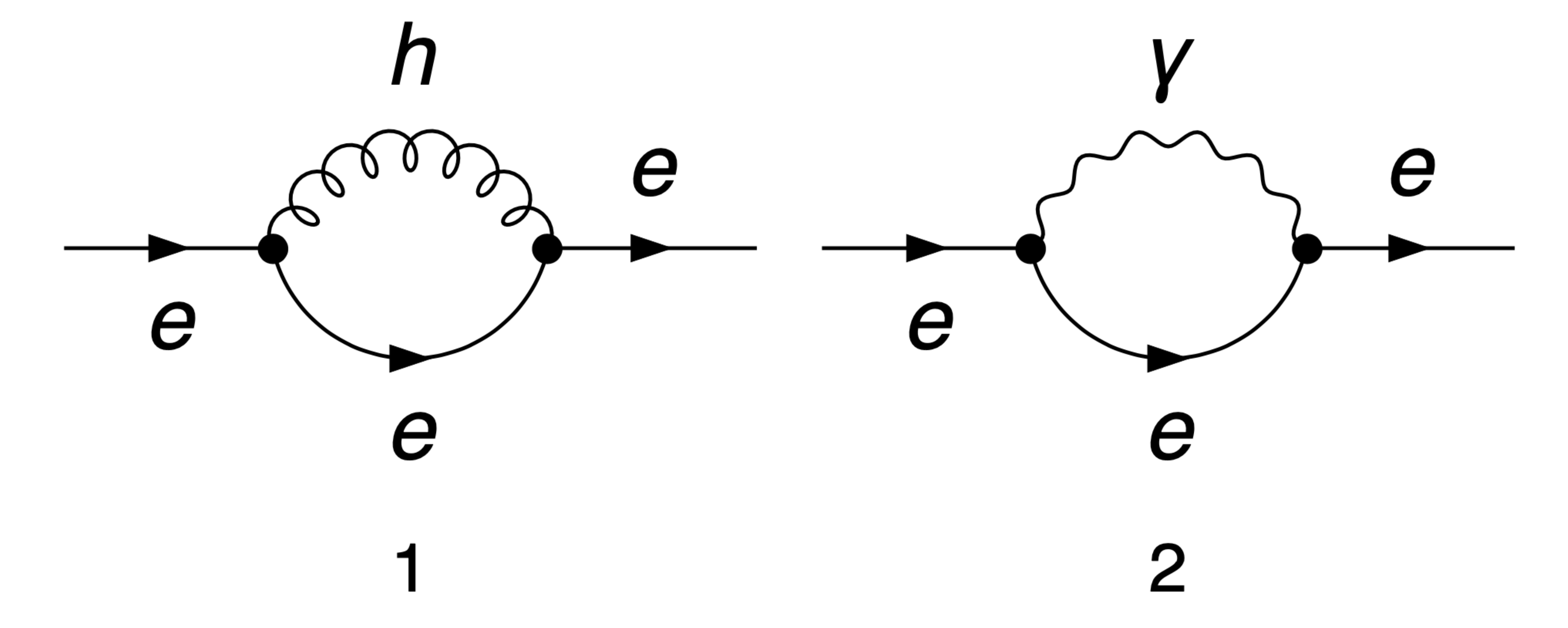}
	\caption{Feynman diagrams for the electron self-energy. Continuous, wavy and wiggly lines represent the electron, photon and graviton propagators, respectively. For the muon self-energy, we just have to change the electron propagators for the muon ones.}
	\label{fig02}
\end{figure}

\begin{figure}[h]
	\includegraphics[angle=0 ,width=8cm]{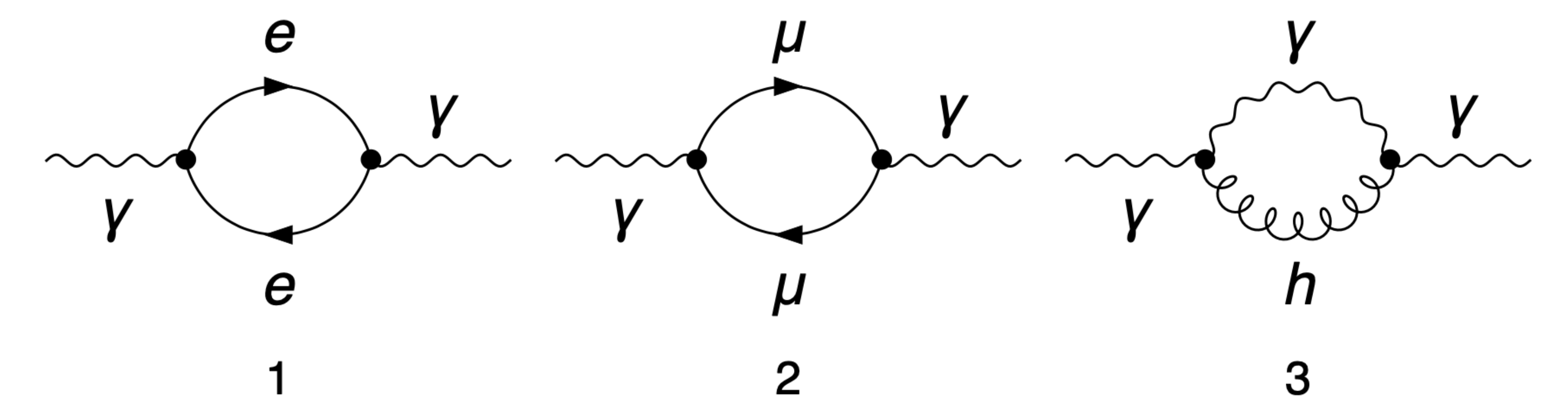}
	\caption{Feynman diagrams for the photon self-energy.}
	\label{fig03}
\end{figure}

\begin{figure}[h!]
	\includegraphics[angle=0 ,width=10cm]{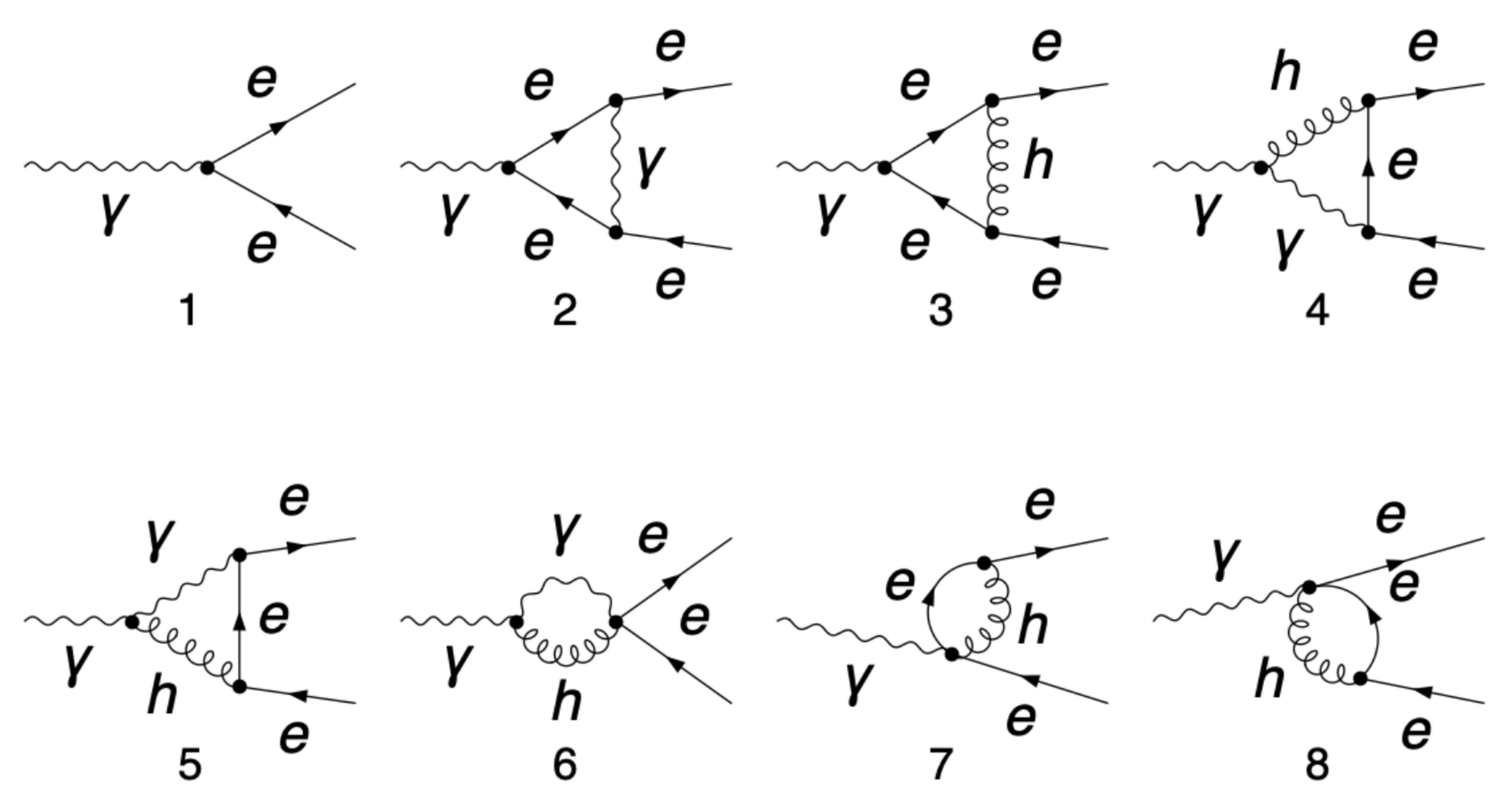}
	\caption{Feynman diagrams to the vertex interaction between electrons and photons up to one-loop order. For the muon vertex interactions, we just have to change the electron propagators for the muon ones.}
	\label{fig04}
\end{figure}

Now, let us calculate at one-loop order the relevant $Z$ factors for the renormalization of the electric charge, so we will consider the one-loop corrections to the self-energy diagrams, depicted in Figs. \ref{fig02} and \ref{fig03}, and to the vertex (see Fig. \ref{fig04}).

We start computing the self-energies of the leptons (electron and muon), Fig.\ref{fig02}. The corresponding expression can be cast as
\begin{eqnarray}\label{1eq02}
-i\Sigma_l(p) &=& \frac{\left(32 \xi_A e^2+37\kappa^2 m_l^2-29\xi_h\kappa^2 m_l^2\right) \slashed{p}-2m_l \left(16 e^2 (\xi_A+3)-\kappa^2 m_l^2(19\xi_h-23)\right)}{512 \pi^2 \epsilon } \nonumber\\
&& + \frac{p^2\kappa^2 \slashed{p}(15\xi_h-19)+12 m_l(3-\xi_h)}{128 \pi^2 \epsilon } \nonumber\\
&&+ \; Z_{2l}^{(1)}~\slashed{p}-m_l Z_{m_l}^{(1)} + \frac{p^2}{M_P^2}\Big(Z_{l-\tilde{e}_1}^{(1)}~\slashed{p}-m_l Z_{l-\tilde{e}_2}^{(1)}\Big) + \mathrm{finite}.
\end{eqnarray}
\noindent The first term of the above equation is renormalized by the counterterms $Z_{2l}^{(1)}$ and $Z_{m_l}^{(1)}$, while the second term is renormalized by the high derivative operators (\ref{ho}).

Imposing to $\Sigma_l(p)$ the renormalization condition, Eqs. \eqref{ren_cond_qed1},\eqref{ren_cond_qed2},\eqref{ren_cond_ho1}, and \eqref{ren_cond_ho2}, we find the following one-loop counterterms:
\begin{subequations}\label{ct01}
\begin{eqnarray}
Z_{2l}^{(1)} &=& -\frac{e^2 \xi_A}{16 \pi ^2 \epsilon }-\frac{(37-29\xi_h) \kappa ^2 m_l^2}{512 \pi ^2 \epsilon }+f_1(m_l,e,\kappa),\label{ct01-1}\\
Z_{m_l}^{(1)} &=& -\frac{3 e^2 m_l}{16 \pi ^2 \epsilon }-\frac{e^2 \xi_A}{16 \pi ^2 \epsilon }-\frac{23 \kappa ^2 m_l^2}{256 \pi ^2 \epsilon }+\frac{19 \kappa ^2 m_l^2 \xi_h }{256 \pi ^2 \epsilon }+f_2(m_l,e,\kappa),\label{ct01-2}
\end{eqnarray}
\end{subequations}
\noindent where the functions $f(m_l,e,\kappa)$ are finite contributions so their specific expressions are not important to our analysis. Choosing the Feynman gauge, $\xi_A=\xi_h=1$, our results are in agreement with those in Ref.~\cite{Bevilaqua:2021uzk}. 

For the photon self-energy, we write the one-loop correction (corresponding to the diagrams in Fig.\ref{fig02}) as
\begin{eqnarray}\label{1eq02}
\Pi^{\mu\nu}(p)=-\left(p^2 \eta^{\mu  \nu }-p^{\mu } p^{\nu }\right)\Pi(p),
\end{eqnarray}
\noindent where
\begin{eqnarray}\label{eq_pi1}
\Pi(p)= Z_3^{(1)}+\tilde{Z}_3^{(1)}\frac{p^2}{M_P^2}+\frac{\left(16 e^2-\kappa^2 p^2(2-3\xi_h)\right)}{96 \pi ^2 \epsilon }+\mathrm{finite}+\mathcal{O}(p^4),
\end{eqnarray}
from which we can see that $Z_3$ is the renormalizing factor for the Maxwell's term, while $\tilde{Z}_3^{(1)}$ renormalizes a higher derivative term like $F^{\mu\nu}\Box F_{\mu\nu}$. Thus, $Z_3$ is the relevant counterterm to the beta function of the electric charge. Notice that the UV divergent part of Eq.(\ref{eq_pi1}) is not dependent on the masses of the leptons.

Imposing the renormalization conditions over $\Pi(p)$, Eqs. (\ref{ren_cond_qed4}) and (\ref{ren_cond_ho4}), we find 
\begin{subequations}\label{ct02}
\begin{eqnarray}
Z_3^{(1)} &=& -\frac{e^2}{6\pi^2\epsilon}+f_3(m_l,e,\kappa),\label{ct02-1}\\
\tilde{Z}_3^{(1)} &=& \frac{\kappa^2 M_P^2(2-3\xi_h)}{96\pi^2\epsilon}+f_4(m_l,e,\kappa),\label{ct02-2}
\end{eqnarray}
\end{subequations}
\noindent where, as in Eq.\eqref{ct01}, $f(m_l,e,\kappa)$ are finite functions whose specific expressions are not important to our analysis. It is important to remark that the counterterm for the Maxwell's term, $Z_3^{(1)}$, is gauge independent.

Contributions to the vertex function up to one-loop order are depicted in Fig. \ref{fig04}. The resulting expression is
\begin{eqnarray}\label{vertex1}
-i\Gamma^{\mu}_l(p) &=& -e\gamma^\mu\left[(1+Z_{1l}^{(1)})+\frac{e^2\xi_A}{16 \pi^2 \epsilon}+\frac{\kappa^2 m_l^2(37-29\xi_h)}{512 \pi^2 \epsilon }\right]\nonumber\\
&&+\tilde{e}(1+Z_{ho4}^{(1)})[\gamma^\mu,\slashed{p}]+\mathcal{O}(p) +\mathrm{finite},
\end{eqnarray}
and, from renormalization conditions (\ref{ren_cond_qed4}) and (\ref{ren_cond_ho4}), we find
\begin{eqnarray}\label{ct031}
Z_{1l}^{(1)} &=&-\frac{e^2 \xi_A}{16 \pi ^2 \epsilon }-\frac{(37-29\xi_h) \kappa ^2 m_l^2}{512 \pi ^2 \epsilon } +\mathrm{finite}
\end{eqnarray}
\noindent ($Z_{ho4}^{(1)}$ absorbs the divergent terms proportional to $\slashed{p}$). Notice that $Z_{1l}^{(1)}=Z_{2l}^{(1)}$, Eq.(\ref{ct01-1}), as required by the Ward identities (see, for instance, Ref. \cite{Srednicki:2007qs}). Setting the Feynman gauge, $\xi_A=\xi_h=1$, our results are in agreement with those in Ref.~\cite{Bevilaqua:2021uzk}. 

We observe from the above result that the vertex counterterm is process dependent in the context of an effective field theory of gravity, since the dependence on $m_l$ implies that we get different counterterms when computed through processes involving electrons or muons, unlike QED in the absence of gravity. However, due to the Ward identity, the universality of the renormalized charge is respected. 

The beta function of the electric charge can now be obtained from the relation between bare and renormalized electric charge, Eq.(\ref{eq_e_0}). As is well known, the Ward identities impose $Z_1=Z_2$ (see, for example, Ref. \cite{Srednicki:2007qs}), so we have $e_0=\mu^{2\epsilon}Z_{3}^{-1/2}e$. Then, $\beta(e)$ at one-loop order can be cast as
\begin{eqnarray}\label{beta_e_qed1l}
\beta(e)=\mu\frac{de}{d\mu}=\frac{e^3}{6\pi^2},
\end{eqnarray}
\noindent which is independent of $\kappa$ and $m_l$, just as in the Einstein-Scalar QED~\cite{Bevilaqua:2015hma}. 

We also expect that the universality of the renormalized charge is also respected at two-loop order. In Ref.~\cite{Bevilaqua:2021uzk}, it was shown that $Z_3$ depends on the mass of the fermionic particle in the Einstein-QED model, so we believe that extending the analysis to the present model, we would obtain $Z_3$ depending on $(m_e^2+m_\mu^2)$. Let us use the fine-structure constant ${\displaystyle \alpha=\frac{e^2}{4\pi}}$ to write $\beta(\alpha)$ as
\begin{eqnarray}\label{beta_e_qed2l}
\beta(\alpha)=\frac{4\alpha^2}{3\pi}\left(1+\frac{5}{4\pi}\frac{M_l^2}{M_P^2}\right)+\frac{\alpha^3}{2\pi^2},
\end{eqnarray}
\noindent where $M_l^2$ is the sum of the squared masses of the leptons. Since the Planck scale is $M_P\approx 1.22\times 10^{19}~\mathrm{GeV}$ and the most massive lepton is the tau ($m_\tau\approx 1.78~\mathrm{GeV}$), the gravitational corrections to the beta function of the fine-structure constant should be of the order of $\sim 10^{-38}$. Even though this effect is completely negligible to the present model, it would become important in the case of the existence of more massive leptons, such as occurs in models based on small extra dimensions~\cite{Grard:2008pr}.

As mentioned in Sec.\ref{sec2}, we could also have computed the renormalized electric charge from the vertex $\frac{e\kappa}{2}(h\eta_{\mu\nu}-h_{\mu\nu})\bar{\psi}\gamma^\mu\psi A^\nu$. The definition of the renormalized fields, $A^\mu\rightarrow Z_3^{1/2}A^\mu$, $\psi\rightarrow Z_2^{1/2}$, and $h^{\mu\nu}\rightarrow Z_h^{1/2}h^{\mu\nu}$, gives us the relation between bare and renormalized coupling constants:
\begin{equation}\label{ct_ek}
 Z_h^{1/2}Z_3^{1/2}Z_2e_0\kappa_0=\mu^{4\epsilon}Z_6e\kappa.
\end{equation}

In order to proceed, we will need to compute the renormalized $\kappa$, which can be done using the vertex with two vector fields and one graviton from \eqref{LA},
\begin{equation}
 \mathcal{L}_A = -\frac{1}{4}F^{\mu\nu}F_{\mu\nu} + \frac{\kappa}{2}h^\tau_{~\nu}F^{\mu\nu}F_{\mu\tau} - \frac{\kappa}{8}hF^{\mu\nu}F_{\mu\nu}.
\end{equation}

The relation between the bare and the renormalized $\kappa$ is given by
\begin{equation}
 Z_h^{1/2}Z_3\kappa_0 = \mu^{2\epsilon}Z_4\kappa.
\end{equation}

\begin{figure}[h!]
	\includegraphics[angle=0 ,width=12cm]{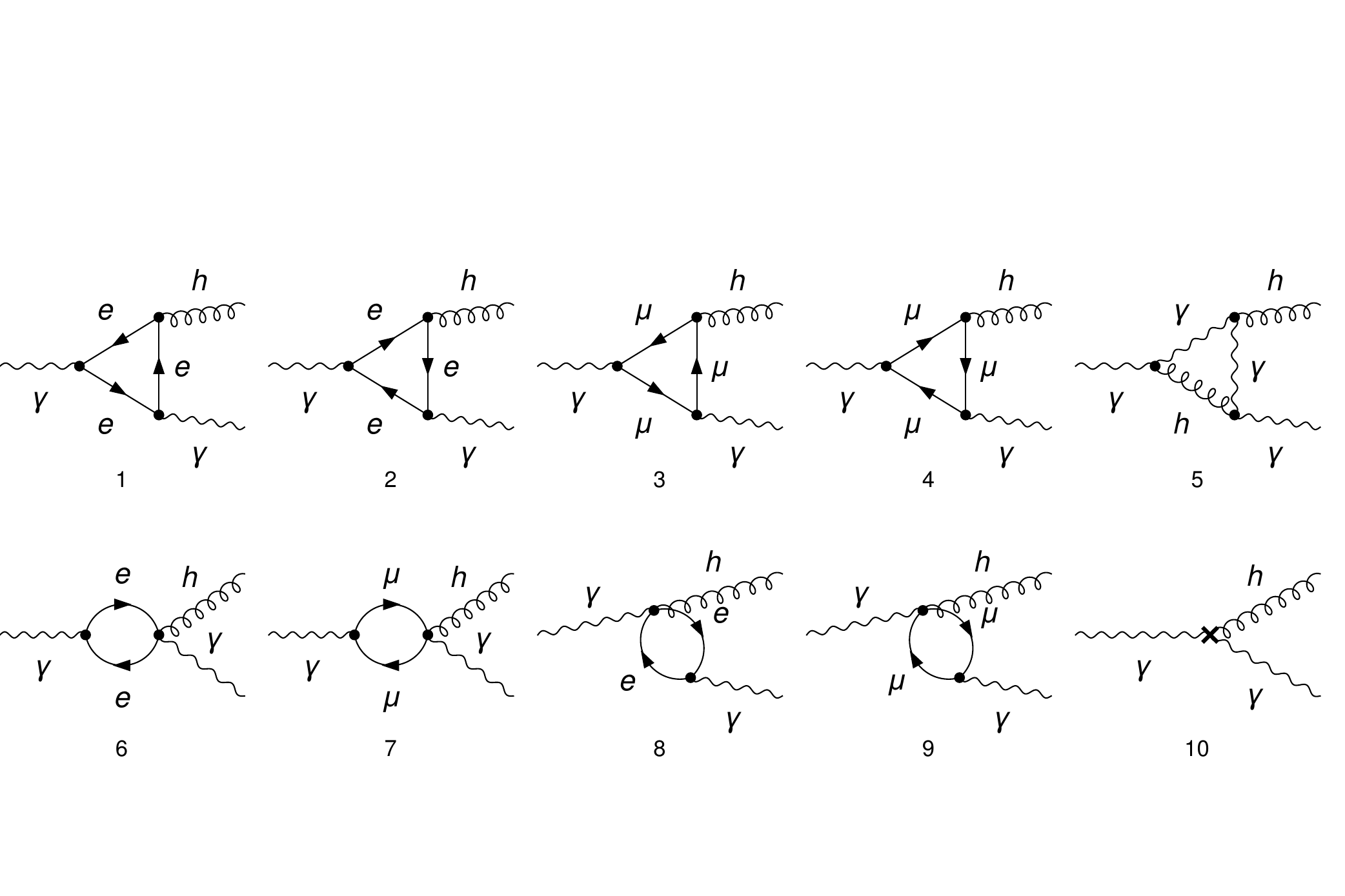}
	\caption{Feynman diagrams to the vertex interaction between photons and a graviton.}
	\label{fig05}
\end{figure}

The diagrams in Fig. \ref{fig05}  allow us to find that $Z_4^{(1)} = -\frac{e^2}{6\pi^2\epsilon} = Z_3^{(1)}$. Therefore, we have
\begin{equation}
 Z_h^{1/2}\kappa_0 = \mu^{2\epsilon}\kappa.
\end{equation}

\begin{figure}[h!]
	\includegraphics[angle=0 ,width=15cm]{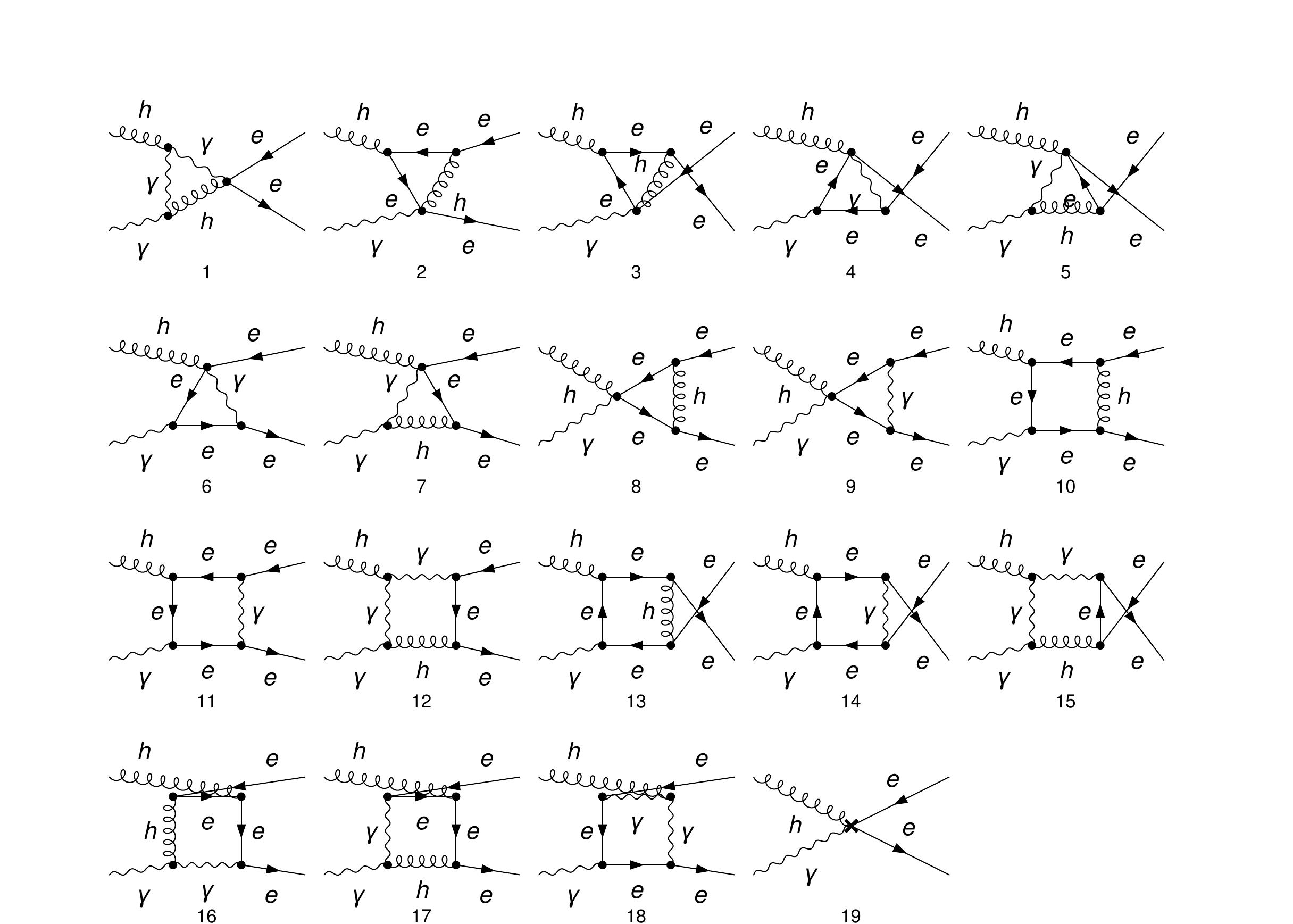}
	\caption{Feynman diagrams to the vertex interaction between fermions, a photon, and a graviton.}
	\label{fig06}
\end{figure}

Using the above equation and $Z_3^{1/2}Z_2e_0 = \mu^{2\epsilon}Z_1e$, we can rewrite \eqref{ct_ek}, at one-loop order, as
\begin{equation}\label{CT_relation}
 \kappa\mu^{2\epsilon}Z_1^{(1)}e\mu^{2\epsilon} = \mu^{4\epsilon}Z_6^{(1)}e\kappa \rightarrow Z_1^{(1)} = Z_6^{(1)},
\end{equation}
and we can confirm this equality by computing the diagrams in Fig. \ref{fig06} to find that
\begin{equation}
 Z_6^{(1)} = -\frac{e^2\xi_A}{16\pi^2\epsilon} + \mathcal{O}(\kappa).
\end{equation}
However, to compute the $O(\kappa)$ terms, we need to consider terms of order $\kappa^2$ in the expanded Lagrangian and we will not pursue this here.

The result in \eqref{CT_relation} is a direct consequence of the gauge invariance of the electrical charge. Although our results suggest that gauge symmetry is preserved, a full analysis on this topic would require a study of the Ward identities of the model, which we expect to do in a future work.

\section{Renormalization via physical processes in Einstein-Scalar QED}

The concept of running coupling constants within effective field theories is in general not as simple as it is for renormalizable field theories. When the theory has a dimensional coupling, the quantum corrections to the scattering amplitude carries a dependence on the energy scale of the process, leading to an operator mixing in the renormalization process \cite{Anber:2010uj}. Therefore, in order to avoid the pitfalls of defining a useful coupling constant, one should consider the renormalization of the model via physical processes. In the present paper we have renormalized n-point functions, rather than scattering amplitudes, and therefore we must now address the question of the validity of this approach.

In previous works, we have used both approaches to discuss the renormalization of the scalar QED coupled to gravity. In \cite{Bevilaqua:2015hma} we have used physical processes and in \cite{Bevilaqua:2021uzk} we have computed the same counterterms using the method we have used here. In this section, we review and analyze the key results from the former, showing that the results are in agreement with the later. 

The model studied in \cite{Bevilaqua:2015hma} is given by
\begin{eqnarray}
 S =&& \int d^4x\sqrt{-g}\Bigr\{\frac{2}{\kappa^2}R - \frac{1}{4}g^{\mu\alpha}g^{\nu\beta}F_{\alpha\beta}F_{\mu\nu} + g^{\mu\nu}(\partial_\mu + ieA_\mu)\phi_j(\partial_\nu-ieA_\nu)\phi_j^\dagger\nonumber\\
 &&-m^2(\phi^\dagger_j\phi_j) - \frac{\lambda}{2}(\phi^\dagger_j\phi_j)^2 + \mathcal{L}_{HO} + \mathcal{L}_{GF} + \mathcal{L}_{CT}\Bigr\},
\end{eqnarray}
\noindent where $j=1,2$. Also,
\begin{eqnarray}
 \mathcal{L}_{HO}=&&\lambda_1\partial^\mu(\phi_1\phi^\dagger_1)\partial_\mu(\phi_2\phi^\dagger_2) + \lambda_2(\phi_1\partial^\mu\phi^\dagger_1-\partial^\mu\phi_1\phi^\dagger_1)(\phi_2\partial_\mu\phi^\dagger_2-\partial_\mu\phi_2\phi^\dagger_2)\nonumber\\
 &&+\lambda_3(\Box\phi_1\phi^\dagger_1\phi_2\phi^\dagger_2+\phi_1\Box\phi^\dagger_1\phi_2\phi^\dagger_2+\phi_1\phi^\dagger_1\Box\phi_2\phi^\dagger_2+\phi_1\phi^\dagger_1\phi_2\Box\phi^\dagger_2) + \cdots.
\end{eqnarray}

The counterterms for the scalar wave function and for the scalar mass were found computing the scalar self-energy and are given by
\begin{subequations}
 \begin{eqnarray}
  &&Z_2^{(1)} = \frac{e^2}{8\pi^2\epsilon}-\frac{m^2\kappa^2}{16\pi^2\epsilon};\\
  &&Z_{m^2}^{(1)} = \frac{3(e^2-\lambda)m^2}{16\pi^2\epsilon},
 \end{eqnarray}
\end{subequations}
while the photon self-energy was used to find
\begin{equation}
 Z_3^{(1)} = -\frac{e^2}{24\pi^2\epsilon}.
\end{equation}

\begin{figure}[h!]
	\includegraphics[angle=0 ,width=6cm]{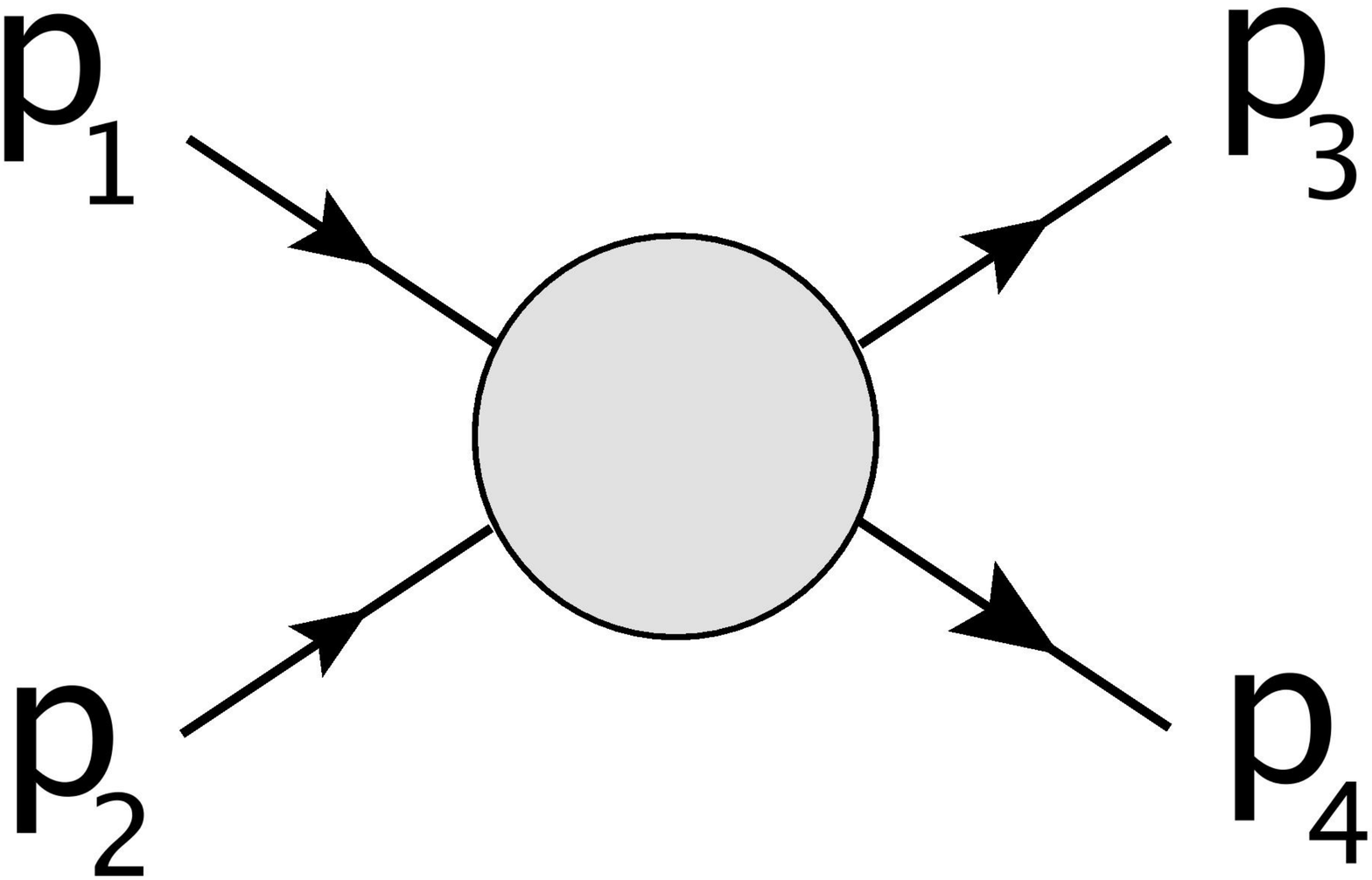}
	\caption{Denomination of external momenta.}
	\label{Mandelstam}
\end{figure}

Our results for the scattering amplitudes found in \cite{Bevilaqua:2015hma} were written in terms of the Mandelstam variables
\begin{eqnarray*}
 && S=(p1+p3)^2=(p2+p4)^2;\nonumber\\
 && T=(p2-p1)^2=(p4-p3)^2;\nonumber\\
 && U=(p4-p1)^2=(p2-p3)^2,
\end{eqnarray*}
and we have also used the constraint $S+T+U = \sum_{i=1}^4 m_i^4$, where $m_i$ are the masses of the external particles (see Fig. \ref{Mandelstam}).

The resulting amplitude for the scattering of two scalar particles can be written as
\begin{eqnarray}
 \mathcal{M} &=& \mathcal{M}_{tree} + \mathcal{M}_{CT} + \mathcal{M}_{1l}\nonumber\\
 && - \lambda +\frac{e^2(S-U)}{T} + \frac{\kappa^2SU}{4T} + \frac{m^2\kappa^2}{2} - \frac{m^4\kappa^2}{2T} + \mathcal{M}_{CT} + \mathcal{M}_{1l},
\end{eqnarray}
\noindent where, for convenience, we have set $\lambda_1=\lambda_2=\lambda_3=0$ at tree level. The expression for the counterterms was found to be
\begin{eqnarray}\label{SQED_CT}
 \mathcal{M}_{CT} =&& -Z_{\lambda}^{(1)} + \frac{2eZ_1^{(1)}(S-U)}{T} + \frac{Z_{\kappa^2}^{(1)}SU}{4T} + \frac{Z_{m^2}^{(1)}\kappa^2}{2} + \frac{m^2Z_{\kappa^2}}{2} - \frac{m^2Z_{m^2}^{(1)}\kappa^2}{T}\nonumber\\
 &&-\frac{m^4Z_{\kappa^2}^{(1)}}{2T} + Z_{\lambda_1}^{(1)}T + Z_{\lambda_2}^{(1)}(S-U)-4m^2Z_{\lambda_3}^{(1)}.
\end{eqnarray}

From the above expression we can see that the electromagnetic counterterm $Z_1^{(1)}$ has a unique kinematic and therefore there is no mixing with higher-order terms. The one-loop scattering amplitude is given by
\begin{eqnarray}
 \mathcal{M}_{1l} =&& \frac{3\lambda^2}{8\pi^2\epsilon} + \frac{3e^4}{8\pi^2\epsilon} - \frac{e^4(S-U)}{4\pi^2\epsilon T} - \frac{11e^2S\kappa^2}{384\pi^2\epsilon} + \frac{e^2m^4\kappa^2}{48\pi^2T\epsilon} + \frac{7e^2m^2S\kappa^2}{484\pi^2T\epsilon} + \frac{5e^2S^2\kappa^2}{384\pi^2T\epsilon}\nonumber\\
 && +\frac{11e^2T\kappa^2}{192\pi^2\epsilon} + \frac{11e^2U\kappa^2}{384\pi^2\epsilon} + \frac{5e^2m^2U\kappa^2}{24\pi^2\epsilon T} - \frac{11e^2SU\kappa^2}{96\pi^2\epsilon T} - \frac{25e^2U^2\kappa^2}{384\pi^2\epsilon T} - \frac{e^2\lambda}{8\pi^2\epsilon}\nonumber\\
 &&-\frac{m^2\kappa^2\lambda}{8\pi^2\epsilon} - \frac{S\kappa^2\lambda}{32\pi^2\epsilon} - \frac{3m^4\kappa^2\lambda}{16\pi^2\epsilon T} - \frac{3T\kappa^2\lambda}{64\pi^2\epsilon} - \frac{U\kappa^2\lambda}{32\pi^2\epsilon} + \text{finite terms}.
\end{eqnarray}

Then, we can compute the counterterms as
\begin{subequations}
 \begin{eqnarray}
  &&Z_1^{(1)} = \frac{e^3}{8\pi^2\epsilon} - \frac{e\kappa^2m^2}{16\pi^2\epsilon};\\
  &&Z_{\kappa^2}^{(1)} = \frac{\kappa^2e^2}{4\pi^2\epsilon};\\
  &&Z_{\lambda_1}^{(1)} = \frac{\kappa^2(\lambda-2e^2)}{64\pi^2\epsilon};\\
  &&Z_{\lambda_2}^{(1)} = \frac{13\kappa^2e^2}{192\pi^2\epsilon}.
 \end{eqnarray}
\end{subequations}

Since $Z_2^{(1)} = Z_1^{(1)}/e$, we have that the Ward identity is satisfied and we only need the photon wave-function counterterm to compute the renormalized electric charge. These results are in agreement with the ones in \cite{Bevilaqua:2021uzk}, which were obtained using the same methods as the present paper. 

We emphasize that this is possible only when each counterterm has its own kinematics, as it is the case for $Z_1$ in \eqref{SQED_CT}. As discussed in \cite{Bevilaqua:2015hma}, the operator mixing occurs when we consider the coupling $\lambda$, since kinematics cannot be used to separate $Z_\lambda$ and $Z_{\lambda_3}$ in \eqref{SQED_CT}.

\begin{figure}[h!]
	\includegraphics[angle=0 ,width=13cm]{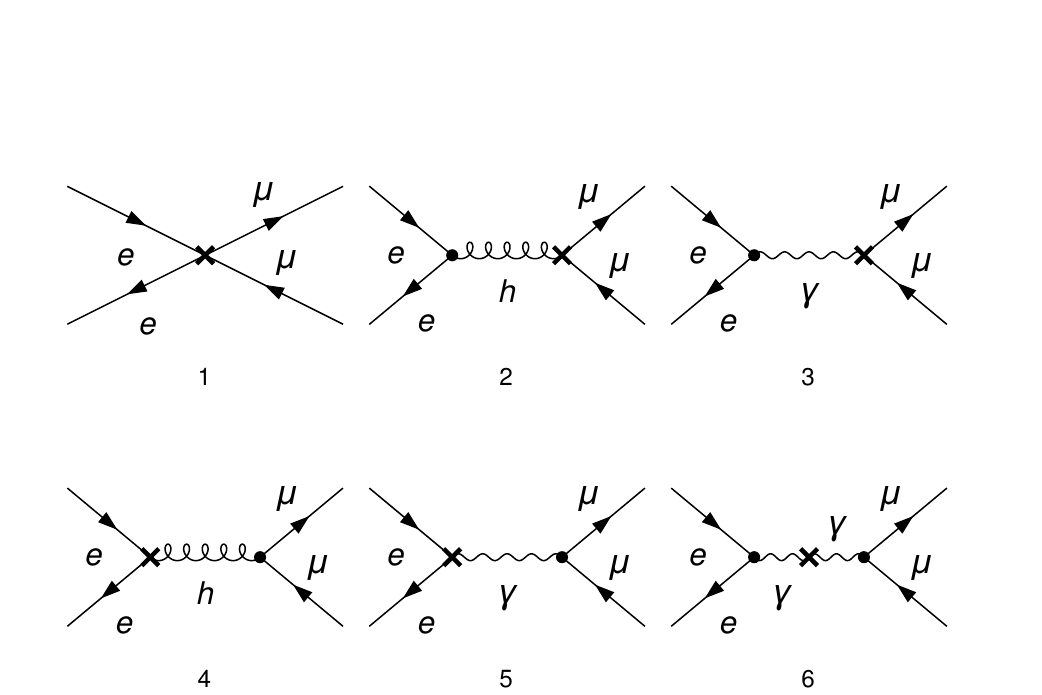}
	\caption{Counterterms to the muon production process.}
	\label{fig07}
\end{figure}

This uniqueness of kinematics happens also for the counterterms in the fermionic case, as can be seen in the muon production process $e^+e^- \rightarrow \mu^+\mu^-$ (see Fig. \ref{fig07}). The amplitude is given by
\begin{eqnarray}\label{Fermions_CT}
 \mathcal{M}_{CT}=&&\frac{\kappa ^2 (Z_5^{(1)}+Z_{5b}^{(1)}-2) \bar{v}(p_1)p_3^\mu u(p2)\bar{u}(p_3)p_{1\mu}v(p4)}{8 S} \nonumber\\
 &&-\frac{\kappa  m_1 m_2 (\kappa  (Z_5^{(1)}+Z_{5b}^{(1)}-2)-2 (Z_{7}^{(1)}+Z_{7b}^{(1)}-2)) (\bar{v}(p_1)u(p_2)\bar{u}(p_3)v(p_4))}{8 S}\nonumber\\
 &&-\Bigr(\frac{\kappa ^2 (T-U) (Z_{5}^{(1)}+Z_{5b}^{(1)}-2)}{32 S}+\frac{i \tilde{e}_4 e (\tilde{Z}_{4}^{(1)}+\tilde{Z}_{4b}^{(1)}) \left(2 m_1^2-2 m_2^2+S\right)}{2 M_P^2S}\nonumber\\
 &&+\frac{16 \pi  \alpha \tilde{e}_3 \tilde{Z}_3^{(1)} \left(m_1^2-m_2^2\right)^2}{M_P^2S^2}+\frac{4 \pi  \alpha \left(4 \tilde{e}_3 m_1^2 \tilde{Z}_3^{(1)}/M_P^2-4 \tilde{e}_3 m_2^2 \tilde{Z}_3^{(1)}/M_P^2+Z_{1}^{(1)}+Z_{1b}^{(1)}-Z_{3}^{(1)}\right)}{S}\nonumber\\
 &&+\frac{4 \pi  \alpha \tilde{e}_3 \tilde{Z}_3^{(1)}}{M_P^2}\Bigr)\bar{v}(p_1)\gamma^\mu u(p_2)\bar{u}(p_3)\gamma_\mu v(p_4),
\end{eqnarray}
\noindent in which $Z_5$ renormalizes the vertex $h\bar{\psi}_1\psi_1$, $Z_7$ the term $m_1h\bar{\psi}_1\psi$, $\tilde{Z}_3$ the $HO$ term $F_{\mu\nu}\Box F^{\mu\nu}$ and $\tilde{Z}_4$ the $HO$ term $F_{\mu\nu}\bar{\psi}_1\gamma^\mu\partial^\nu\psi$, while the counterterms with subscript $b$ renormalize the terms with $\psi_2$. As we can see from \eqref{Fermions_CT}, the counterterms have different kinematics and the ones that have the same will renormalize different processes. Therefore, we expect that the results obtained from the computation via physical processes will be consistent with the ones we obtained in this paper. However the proper analysis of the subject for the Einstein-QED system will be left for a future work, since the full calculation is very subtle and rich in details.

\section{Final Remarks}\label{summary}

In summary, we have computed the gravitational corrections to the one-loop n-point functions to find the renormalization constants in quantum electrodynamics, with an arbitrary gauge, showing that the electric charge can be universally defined to the interaction between photons and leptons. Our calculations show that although the renormalization of the three-point functions are process dependent, leading to different vertex counterterms, the Ward identity ensures the universality to the renormalized electric charge. We have also argued that our results would stand if physical processes were used to renormalize the model.

It is important to highlight that even though the renormalization of the electric charge does not receive any one-loop gravitational correction, the effect of the metric fluctuations appears in the renormalization of other free parameters of the theory, such as the fermion wave-function renormalization constant (Fig. \ref{fig02}), the vertex function renormalization constant (Fig. \ref{fig04}) and in the renormalization constants of high order operators Eq.\eqref{ho}.

\section*{ACKNOWLEDGMENTS}
The authors would like to thank J. Gegelia for useful comments and the referee for the critical reading and their suggestions, leading to substantial improvements of the manuscript. Huan Souza is partially supported by Coordena\c{c}\~ao de Aperfei\c{c}oamento de Pessoal de N\'ivel Superior (CAPES).

\newpage

\end{document}